\documentclass{aa}

\input{epsf}
\def\lesssim{\mathrel{\hbox{\rlap{\hbox{\lower4pt\hbox{$\sim$}}}\hbox{$<$}}}}
\begin{document}

   \thesaurus{03     
              (11.01.2; 
		11.14.1; 
		11.17.3; 
		13.21.1)} 
\title{Tidal Disruption Eddington Envelopes around Massive Black Holes}

\author{Andrew Ulmer\inst{1}, Bohdan Paczy\'nski\inst{2},
and Jeremy Goodman\inst{2}}
\institute{Max-Plank-Institut f\"ur Astrophysik, Karl-Schwarzschild-Str. 1,
85740 Garching, Germany
\and
Princeton University Observatory, Peyton Hall, Princeton, NJ 08544}

   \offprints{A. Ulmer}

   \date{Submitted November 18, 1997}

\titlerunning{Tidal Disruption Eddington Envelopes}
\authorrunning{Ulmer, Paczy\'nski, \& Goodman}
\maketitle

   \begin{abstract}
Optically-thick envelopes may form following the tidal
disruption of a star by a massive black hole. 
Such envelopes would reprocess hard radiation from accretion close
to the black hole into the UV and optical bands producing
AGN-luminosity flares with
duration $\sim 1$~year.
We show that due to relativistic effects, the envelopes are convective.
If convection is efficient, then the
structure of the envelopes is similar to that described in previous work;
however, the photospheric radius is shown to be very sensitive
to the luminosity at the envelope base, suggesting that either the envelope
collapses or the envelope expands to a maximum radius at which point a wind
may set in.
For an envelope without winds, we find a maximum photospheric radius of
$\sim 10^{16}$~cm (i.e. minimum effective
temperature $\sim 6,000$~K).
The evolution of the envelopes is described based on simple
energy arguments.

   \end{abstract}

%

\section{Introduction}

One likely outcome of the tidal disruption of a star by a massive black hole
is a bright flare with duration of a few months to years (e.g. Rees 1988,
Ulmer 1998a; hereafter U98).
The flares are generally thought to be quite hot.
For example, the temperature associated with an Eddington luminosity
emitted from a spherical photosphere at the tidal radius is
\begin{eqnarray}
\label{teff1}
T_{\rm eff} &\approx& \left( { L_{\rm E} \over 4 \pi R^2_{\rm t} \sigma}
\right)^{1/4} \\
 &\approx& 2.5 \times 10^{5} M_6^{1/12} \left( {R_\star \over R_\odot}
\right)^{-1/2} \left( {M_\star \over M_\odot}
\right)^{-1/6} ~{\rm K},
\end{eqnarray}
where $M_6$ is the black hole mass in units of a million solar masses
and $M_\star$ and $R_\star$ are the mass and radius of the star.
Consequently, the discussions of the spectra and observability of
flares have focused on 
hard emission (e.g. Sembay \& West, 1993) or the extreme, $\sim 7.5$~magnitude,
bolometric corrections to optical (e.g. U98).
Loeb \& Ulmer (1997, hereafter LU97) make the interesting
suggestion that if part of the tidal debris forms an extended envelope
around the black hole, then the light may be largely reprocessed
down to optical with effective temperatures of
\begin{equation}
T_{\rm eff}\approx 1.3\times 10^4 \left({M_{\rm BH}\over
10^6 M_\star}\right)^{1/4} ~{\rm K}.
\label{eq:teff}
\end{equation}
In the model, an accretion disk forms close to the black hole, and 
the energy released from the disk is reprocessed by the extended
envelope which connects in an unknown way to the disk.

In Sect.~\ref{clossec}, we show that without knowledge of the base
luminosity which is provided by disk accretion onto the black hole,
it is impossible to specify uniquely the resulting envelope structure.
In Sect.~\ref{schsec}, we discuss the small, but important, redshift effects
from the Schwarzschild geometry and show that the envelopes are convective.
The system's time scales are investigated in Sect.~\ref{evtime}, and
using simple energy arguments, we make predictions
regarding the evolution of the envelope.
A discussion of these results is presented in Sect.~\ref{efinsec}.

\section{Lack of Closure Relations  \label{clossec}}

The envelope density profile derived in LU97 (Eq.~3)
results in a logarithmicly diverging mass, so
an artificial cutoff of the atmosphere was taken in LU97 (Eq.~8).
A related problem is that the pressure close to the photosphere
is not handled in a self-consistent manner. In this section, we address
these problems and discuss their implications.

We operate in a Newtonian potential, assume
that the envelope is radiative, and link the envelope model to
the atmosphere using the Eddington approximation
in which the surface temperature, $T_{\rm 0} = (1/2)^{1/4} T_{\rm eff}$,
where $T_{\rm eff}$ is the effective temperature.
At the photospheric radius, $r_{\rm phot}$, where the
optical depth is 2/3, $T=T_{\rm eff}$. The Eddington approximation
requires that the atmosphere be thin so that the surface is not much
larger than $r_{\rm phot}$.
The hydrostatic equation is:
\begin{equation}
\label{eq2}
{d P \over d r} = - {GM \rho \over r^2}
\end{equation}
where $M$, is the mass of the black hole alone.
We neglect the envelope mass since
$M_{\rm env}/M_{\rm BH} \sim 10^{-6}$.
The equation of radiative transfer is
\begin{equation}
\label{eq3}
{d P_{\rm rad} \over d r} =  - { \kappa L \rho \over 4 \pi c r^2},
\end{equation}
where the luminosity, $L$, and absorption coefficient, $\kappa$, are
assumed to be constant throughout the envelope.
Dividing the previous two equations yields the familiar result:
\begin{equation}
\label{eq4}
1- {d P_{\rm gas} \over d P} =
{d P_{\rm rad} \over d P} = {\kappa L \over 4\pi c G M} =
{L \over L_{\rm E}},
\end{equation}
where $L_{\rm E}$ is the Eddington luminosity.
The right side of equation~\ref{eq4} is constant, because
the mass is dominated by the mass of the black hole,
the luminosity is wholly supplied by the accretion disk at the
base of the envelope, and $\kappa$ is dominated by electron scattering.
Integrating, we find
\begin{equation}
\label{eq5}
P_{\rm gas} = (P-P_{\rm 0}) \left( 1-  {L \over L_{\rm E} }\right),
\end{equation}
where the surface pressure, $P_{\rm 0}$, is, in the Eddington approximation,
equal to $P_{\rm rad,0} = L/6 \pi c r_{\rm phot}^2$.
This surface term was omitted in LU97.

Because pressure goes as the fourth power of temperature, throughout most
of the envelope the surface pressure term in Eq.~\ref{eq5} is unimportant, and
\begin{equation}
\label{eq6}
P_{\rm gas} = P \left( 1-  {L \over L_{\rm E} }\right) = P \beta,
\end{equation}
where $\beta$ is a constant.
It then follows from the equation of state of the gas
($P_{\rm gas}=\rho k_{\rm B} T/ \mu m_{\rm h}$) and the radiation
($P_{\rm rad}=a T^4/3$) that everywhere in the envelope except
at low optical depth,
\begin{equation}
\label{eq7}
P = A\rho^{4/3} = \left({  k_{\rm B} \over \beta
\mu m_{\rm h} }\right)^{4/3}
\left( { 3 (1- \beta) \over a }\right)^{1/3} \rho^{4/3}.
\end{equation}
This relation was assumed in LU97.

It is now possible to solve the hydrostatic equation (Eq. \ref{eq2}) for
temperature or, equivalently, density by using the approximate
relation between $P$ and $\rho$ (Eq.~\ref{eq7}):
\begin{eqnarray}
\label{eq8}
T &=& {\beta \mu m_{\rm H} GM \over 4 k_{\rm B} }
\left( {1\over r} - {1\over r_0} \right) \\
\rho &= & \left( {GM \over 4 A} \right)^3 \left( {1 \over r} -
{1\over r_0} \right)^3,
\label{deneq}
\end{eqnarray}
where $r_0$ is a constant of integration which was neglected in LU97.
The constant is determined by the requirement that at the photospheric
radius, $r_{\rm phot}$, $T=T_{\rm eff}$. 

Following LU97, we find a relationship between the envelope mass and
$\beta$. The envelope mass is found by integrating
Eq.~\ref{deneq} from the base radius,
\begin{equation}
r_{\rm b} \sim R_{\rm T} = R_\odot (M_{\rm BH} /M_\odot)^{1/3} \sim
10^{13}~{\rm cm}
\end{equation}
to $r_{\rm phot}$
(The exact location of base radius is an unknown; it is the poorly understood
interface between the disk and envelope. The base radius, as argued in
LU97, should be close to the tidal radius, because that is where
the dynamical effects of angular momentum are expected to become important.)
\begin{eqnarray}
M_{\rm env} &=& \int_{r_{\rm b}}^{r_{\rm phot}} \rho 4 \pi r^2 dr \\
&\approx&
4 \pi \left( {GM \over 4 A} \right)^3 \left[ \ln(r_{\rm phot}/r_{\rm b})
- 1.8 \right] \\
 &\approx& 1.9 \times 10^{15} M_6^3
\beta^{4} 
\left[ \ln(r_{\rm phot}/r_{\rm b}) -1.8\right]  M_\odot.
\label{expo}
\end{eqnarray}
An envelope which contains no more than a solar mass
must be extremely close to the Eddington limit with $\beta \lesssim 10^{-4}$.
Eq.~\ref{expo} shows that the photospheric radius is exponentially
sensitive to $\beta$, suggesting that very small changes in base luminosity,
which is controlled by a poorly understood accretion in the disk, may
create large changes in the photospheric radius.
For example, in this Newtonian approximation, the steady state
radius would have to change from $10^{14}$ to $10^{15}$~cm when
$(1-L / L_{\rm E})$ changes from $\sim 8 \times 10^{-5}$
to $\sim 5 \times 10^{-5}$.
This result illustrates the extreme fine tuning
(to three parts in $10^{5}$) which was required
in LU97 to reach the steady state solution. As we describe
in Sect.~\ref{schsec}, the fine tuning required in the Schwarzschild case
is less, but the level is $\sim 2\%$ {\em above} the local critical
luminosity.

We now estimate the maximum radius at which the Eddington
approximation is valid. The approximation
requires that the atmosphere be thin, or equivalently,
that the gas pressure scale height be much less than
the radius at the photosphere.
This limit is also a physical dividing line,
because for further extended atmospheres,
one expects high mass loss as the atmosphere is nearly isothermal at
$\tau \ll 1$, so the escape velocity falls faster than the thermal velocity.

The gas pressure scale height at the photosphere can be evaluated
using Eqs.~\ref{eq2} and \ref{eq6} as:
\begin{eqnarray}
h_{\rm Pg}^{-1} & = & - {1\over P_{\rm gas}} {d P_{\rm gas} \over d r} \\
{h_{\rm Pg} \over r} & = &  {k_{\rm B}
(L_{\rm E}/4 \pi \sigma r_{\rm phot}^2)^{1/4}
\over \mu m_{\rm h} } { r_{\rm phot} \over  \beta G M  }
\end{eqnarray}

At $r_{\rm phot}$, the
gas pressure scale height must be much less than the photospheric radius:
\begin{eqnarray}
1 \gg {h_{\rm Pg} \over r_{\rm phot}}& = & {k_{\rm B}
(L_{\rm E}/4 \pi \sigma)^{1/4}
\over \mu m_{\rm h} }{ r_{\rm phot}^{1/2} \over  \beta G M  } \\
r_{\rm phot} & \ll &  1.5  \times 10^{16}
\left({M_{\rm env}\over 0.5 M_\odot}\right)^{1/2} M_6^0~{\rm cm}.
\end{eqnarray}
where we have written
$\beta$ in terms of the envelope mass (Eq.~\ref{expo}).
Note that result is independent of the black hole mass.
The effective temperature therefore has the same scaling as in LU97:
\begin{equation}
T_{\rm eff} \gg 4,500 \left( {M_{\rm BH} \over M_{\rm env}}
\right)^{1/4} ~{\rm K}
\end{equation}
These limiting results are surprisingly close to those of LU97 who found
$r_{\rm phot} \sim 2 \times 10^{15}  ~{\rm cm}$ and
$T_{\rm eff} \sim 15,000~{\rm K}$.
The difference, of course,
is that our results show that smaller photospheric radii could exist.
The exact photospheric radius cannot be determined without knowledge
of the inner luminosity source.

An additional way
to see the wind constraint on the radius is to consider
$r_0^{-1}$ in Eq.~\ref{eq8}, which enforces the condition that
$T = T_{\rm eff}$ at the photosphere (which radiates at the Eddington limit).
The constant is zero when $r_{\rm phot} \sim 10^{15}$, and becomes negative
for larger values of $r_{\rm phot}$. Negative values of $r_0^{-1}$ give
(see Eq.~\ref{eq8}) regions at the top of the envelope which approach
an isothermal state and which would likely drive winds.

Even before the radius expands so far that the
pressure scale height becomes comparable with the radius, winds
could begin to play an important role--either in altering
the envelope structure or in removing much the envelope mass.
A estimate of the importance of the winds may be found
by connecting, at the photospheric radius, isothermal wind solutions to
the envelope solutions. 
Because of the extremely low densities
($\sim 10^{-15}$g/cm$^3$), 
the envelope is ionized and the cross-section
is largely provided by electron-scattering, so an isothermal wind
powered by the continuum cross-section (rather than line transitions),
may be most appropriate.
In this case, the following equation describes the outflow
(e.g. Kudritzki, 1988):
\begin{equation}
(1- {v_{\rm s}^2 \over v^2} ) v {  d v \over   d r } =
{2 v_{\rm s}^2 \over r} - {G M_{\rm BH} \over r^2 }
(1- {L \over L_{\rm E} } ),
\label{wind1}
\end{equation}
where $v_{\rm s}$ is the sound speed (in our case, isothermal so that
$v_{\rm s} = (k_{\rm B} T_{\rm eff} / \mu m_{\rm h})^{1/2}$ and
the sonic point, $r_{\rm s} = GM (1-L/L_{\rm E})/2 v_{\rm s}^2$.
Dimensionless radius, $\eta = r /r_{\rm s}$,
and dimensionless velocity, $u = v / v_{\rm s}$,
allow Eq.~\ref{wind1} to be neatly integrated to obtain:
\begin{equation}
\ln u - {u^2 \over 2} = - {2\over \eta} - 2 \ln \eta + 1.5.
\end{equation}
Connecting such outflows to the Eddington envelopes allows for determination
of mass loss as a function of envelope photospheric radius.
Generally, the time for such a wind to significantly reduce the envelope mass
is at least 1000~years---many times longer than the other relevant time scales
for the problem (results are shown in Fig.~1). For photospheric radii
larger than $\sim 3 \times 10^{15}$~cm, the sonic point occurs below the
photosphere, showing, in agreement with the
limiting radius found above, that the hydrostatic solution
(Eqs.~\ref{eq8}~and~\ref{deneq}) is no longer valid.

\section{Effects in the Schwarzschild Metric \label{schsec}}

The Schwarzschild geometry has a significant impact on
envelopes which radiate near the Eddington limit.
At first appearance, the Schwarzschild geometry
seems to be an unimportant correction to the problem, because the
base of the photosphere is at
$\sim 25 R_{\rm S}$, and redshift effects are of order 2\%,
$(1+z) \sim (1+R_{\rm S}/ 2 r)$. However, because the
envelopes are very close to the Eddington limit, as is required to
produce an extended photosphere,
the envelope structure is very sensitive to $\beta =  1-L/L_{\rm E}$.
In particular for envelopes of $\sim 1 M_\odot$, $\beta$
is $\sim 10^{-4}$, so a 2\% reduction in luminosity will
significantly alter the envelope structure.

We quote below a number of results which were obtained in the study of
x-ray bursts and
envelopes around neutron stars (Paczy\'nski \& Anderson 1986;
Paczy\'nski \& Pr\'oszy\'nski 1986).
For our purposes, the most important feature of the relativistic
stellar structure equations is that the local critical luminosity does not
scale with radius in the same manner as the local luminosity with the
consequence that the structure is convective rather than radiative.
Specifically, the local luminosity and
critical luminosity determined by the local gravitational
forces (equivalent to the Eddington luminosity at large r)  are
\begin{eqnarray}
\label{leq1}
L&=& L_{\infty} \left( 1 - { R_{\rm S} \over r} \right)^{-1} \\
\label{leq2}
L_{\rm cr}&=& {4 \pi c GM \over \kappa}
\left( 1 - {R_{\rm S} \over r} \right)^{-1/2}, 
\end{eqnarray}
where $L_\infty$ is the luminosity as measured far from the black hole.
Because of the different scalings,
a near critical luminosity at the surface requires a super-critical
luminosity at the base. As a consequence the star is convective.
Alternatively, the convective nature can be seen by comparing the
radiative and adiabatic gradients:
\begin{eqnarray}
\nabla_{\rm ad} &=&
{1 \over 4} \left[{1 - {3\over 4} \beta
\over 1 -{3\over 4} \beta - {3 \over 32} \beta^2}\right]
\approx {1 \over 4} \left[ {1 + {3\over 32}  \beta^2}\right]
\\
\nabla_{\rm rad} &=&
{1 \over 4} \left[{L_\infty \over (1-\beta) L_{\rm E}}
\left( 1 - {R_{\rm S} \over r} \right)^{-1/2} + {4 P\over \rho_0 c^2}
\right] \times \tilde{P}\\
 & \approx &
{1 \over 4} \left[{L_\infty \over (1-\beta) L_{\rm E}}
\left( 1 - {R_{\rm S} \over r} \right)^{-1/2} \right]\\
\tilde{P} &=&\left[ 1+  {(4- 1.5 \beta) P\over \rho_0 c^2} \right]^{-1}.
\end{eqnarray}
Even for a near critical surface luminosity, $\nabla_{\rm rad}$ quickly
becomes larger than $\nabla_{\rm ad}$ as the coordinate, $r$, decreases.

As a consequence,
the envelope must be convective. Whether or not the convection
is efficient is difficult to determine because in radiation
dominated regimes,
convection is not fully understood, although progress is being made
(e.g., Arons 1992).
If convection is efficient, then by definition, $\nabla_{\rm ad}$ is nearly
a constant.
The equation of the temperature gradient yields:
\begin{equation}
{d P_{\rm rad} \over d P} = 
{1 \over 4} \left[{1 - {3\over 4} \beta
\over 1 -{3\over 4} \beta - {3 \over 32} \beta^2}\right]
\approx {1 \over 4} \left[ {1 + {3\over 32}  \beta^2}\right]
\end{equation}
so $P_{\rm rad}/P$ is very closely a constant (to $\sim 1$
part in $10^{-9}$), and we recover the polytropic
equation of state $P=A \rho^{4/3}$, which was used in the previous section.
Assuming that convection is efficient, we can determine the relationship
between the photospheric radius, the envelope mass, and the surface
luminosity in the same manner as in Sect.~\ref{clossec}.
Ignoring the relativistic terms in the hydrostatic stellar structure equations,
which is a good approximation (to $\sim 2\%$)
because the terms enter multiplicatively rather than as differences,
we recover eqn.~\ref{expo}.
In contrast to the Newtonian case, in the Schwarzschild case, 
the luminosity at the base must be locally super-critical in order
to support an extended envelope with a near-Eddington surface luminosity.

If the luminosity is $\sim 2\%$ super-critical, and the envelope
is able to expand, then photospheric radius is still sensitive to
the base luminosity, but not exponentially sensitive as is the case for
a Newtonian atmosphere. The luminosity at the photosphere is nearly
equal (to better than one part in $10^4$) to the critical luminosity
(Eqs.~\ref{leq2}), so
\begin{equation}
L_{\rm phot} \approx L_{\rm cr} = 
L_{\rm E} \left( 1 - {R_{\rm S} \over r_{\rm phot}} \right)^{-1/2}. 
\end{equation}
Using equation~\ref{leq1} to write the luminosity at the base, $L_{\rm b}$
as a function of photospheric radius, yields the result:
\begin{equation}
r_{\rm phot}  \approx {R_{\rm S} \over  1- (L_{\rm E}/L_{\rm b})^2}.
\end{equation}
In steady state the photospheric radius would increase from $10^{14}$
to $10^{15}$~cm with a $0.1\%$ change in base luminosity.

There may be additional effects which compete with the relativistic effects.
For neutron stars, the temperatures at the base of the envelope are so
high that relativistic corrections to the Thompson cross section become
important (e.g. Paczy\'nski \& Anderson 1986), but the temperatures around
the tidal disruption created envelopes never reach such high temperatures.
In our case, slight rotation may serve to reduce the critical luminosity
at small radii, in the plane of rotation. Along the rotation axis, the
envelope would likely be convective for the reasons described above.

\section{Evolution in Time \label{evtime}}

Both the Newtonian and relativistic envelopes require fine tuning
of the luminosity in order to produce static extended envelopes, because
$r_{\rm phot}$ depends sensitively on the luminosity.
In the Schwarzschild case,
unless the base of the envelope knows to shine locally at
$\sim R_{\rm S}/2 r_{\rm b} \sim 2\%$ above the local critical luminosity,
and the luminosity is restricted to a narrow regime, it appears
difficult to maintain the type of envelopes discussed in Sect.~\ref{clossec}.
The dependence on luminosity is strong; a fraction of a percent
change in base luminosity could increase the steady-state radius from
$10^{14}$ to $10^{15}$~cm.
Since the energy source (the accreting torus) is probably not extremely
well-coupled to the envelope, it is unlikely that the base luminosity
is so finely tuned. Therefore, we consider how the photosphere will
change with time for different base luminosities.

We identify three cases. If the base luminosity is sub-critical,
the envelope is not much bigger than the accreting torus.
When the luminosity is super-critical, the envelope must expand.
If the super-criticality is less than $\sim 2\%$, the expansion is modest
and the envelope may remain in hydrostatic and thermal equilibrium.
If the base luminosity is super-critical by a significant factor, e.g. 2,
then the envelope expands hydrostatically on a thermal time scale, until
it expands so much that the dynamical time scale becomes shorter than the
thermal time scale. After that, the outer parts of the envelope are no
longer in hydrostatic equilibrium, but this does not imply that they are
instantly lost. The envelope may continue to expand, perhaps even
with a structure similar to the hydrostatic solution,
until the scale height of atmosphere becomes comparable to the radius
(as discussed in Sect.~\ref{clossec}) at which point a wind will set in.

   \begin{figure}
\hspace{0cm}\epsfxsize=8.8cm \epsfbox{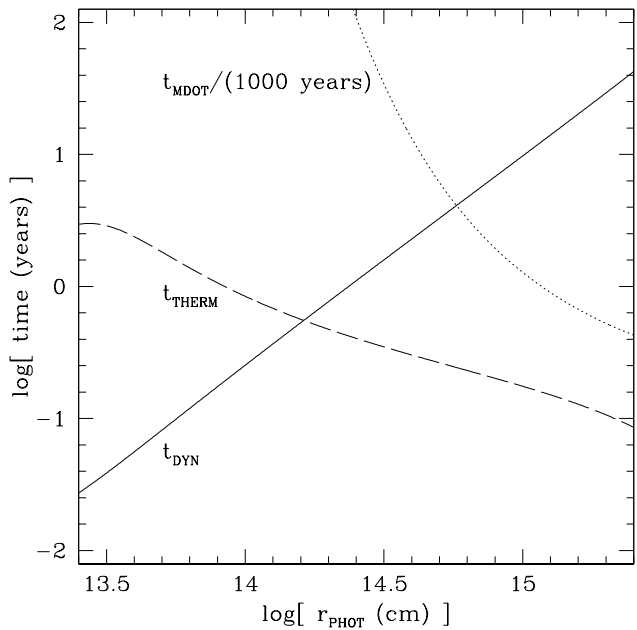}
      \caption[]{Time
scales for envelopes as a function of photospheric radius:
	dynamical time, thermal time, and time for wind driven mass loss.
The thermal time scale is calculated assuming a energy input of
$2 L_{\rm E}$ and energy output of $L_{\rm E}$ and $r_{\rm b}=
10^{13}$~cm.
The envelopes can stay in hydrostatic equilibrium only as long
as $t_{\rm dyn} < t_{\rm therm}$. The mass loss to winds produced
by an isothermal outflow are small enough that the do not appear to
affect the system.
              }
   \end{figure}

We now examine the expansion of the envelope.
The relevant time scales are the dynamical time scale, the thermal
time scale, and the radiation time scale, over which time all matter
would be accreted.
The first two time scales are shown in Fig.~1 as a function
of photospheric radius.

The dynamical time scale grows with photospheric radius:
\begin{eqnarray}
t_{\rm dyn} &=& \left(G M_{\rm BH} (1- L/L_{\rm E}) \over
r_{\rm phot}^3 \right)^{-1/2} \\
&\approx& 13 r_{15}^3 M_6^{-1/2}~{\rm years}.
\end{eqnarray}

The thermal time scale, defined as the time scale to unbind the
envelope assuming the energy in the envelope increases at a rate
of $L_{+} = L_{\rm in}- L_{\rm out} \approx L_{\rm E}$
(e.g. if the base luminosity were $2 L_{\rm E}$
and the surface radiated at $L_{\rm E}$).
The total energy of the hydrostatic envelope is
$E_{\rm tot} = E_{\rm grav} + E_{\rm internal},$
where the internal energy is
\begin{eqnarray}
E_{\rm internal} &\approx& \int_{r_{\rm b}}^{r_{\rm phot}}
4 \pi r^2 P_{\rm rad} {\rm d}r\\
&\approx& {3 \over 4} {G M_{\rm BH}
M_{\rm env} \over r_{\rm b} [ \ln(r_{\rm phot}/r_{\rm b}) -1.8]}.
\end{eqnarray}
Note that the energy is strongly dependent on the location of the inner
boundary, $r_{\rm b}$.
The gravitational energy, $E_{\rm grav} \approx -(4/3)E_{\rm internal}$,
so that the
thermal time scale is
\begin{equation}
t_{\rm therm} = {E_{\rm internal} \over 3 L_{+} } \approx 0.2~
\left( {10^{13}~{\rm cm} \over r_{\rm b}} \right)
\left({L_{\rm E} \over L_{\rm +}}\right)~{\rm years},
\end{equation}
with a complex dependence on photospheric radius and black hole mass.
The thermal time scale depends both on the net energy injection rate and
on the base radius of the envelope. As both parameters depend on the
unknown physics of the torus, and as such, their exact values are
uncertain. We believe the base radius to be near the
tidal radius $\sim 10^{13}$~cm, where rotational support likely becomes
important. Similarly, thick tori may produce luminosities up to a few times
the Eddington limit, but the exact value cannot be predicted.
These parameters enter
into the thermal time scale multiplicatively, and together may lengthen
or shorten the thermal time scale by a factor of 2 or 3.

Lastly, the radiation time is the time to radiate, at the Eddington limit,
all of the accretion energy of the tidal debris:
\begin{equation}
t_{\rm rad} \approx 20 M_6^{-1}~{\rm years}, 
\end{equation}
for an accretion efficiency of $10\%$ and an envelope mass of $0.5 M_\odot$.
This time scale does not depend on the photospheric radius, and is generally
longer than the other relevant time scales of interest (for a $10^6 M_\odot$
black hole).

\begin{figure}
\hspace{0cm}\epsfxsize=8.8cm \epsfbox{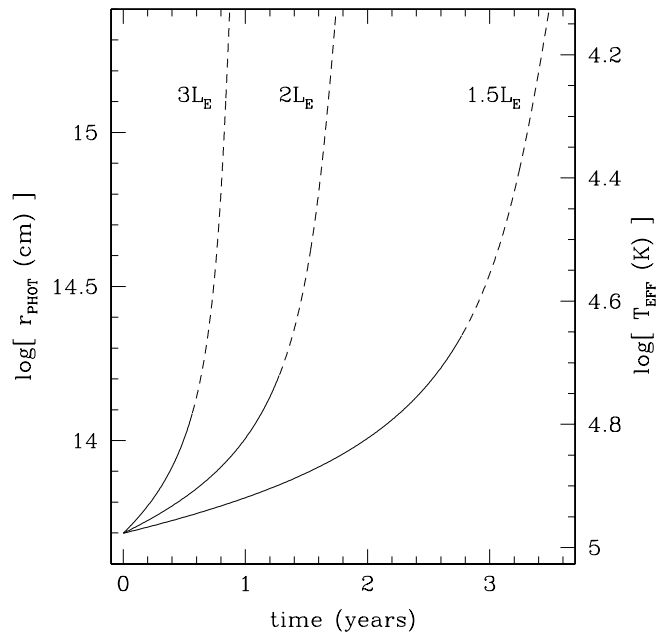}
      \caption[]{Radius and temperature for sequences
of expanding envelopes labeled with three 
different energy input rates (each envelope radiates at the Eddington limit)
and a base radius of $10^{13}$~cm. The solid lines are the phase in
which the envelopes are in hydrostatic equilibrium. The dashed lines
signify that the envelope is not in hydrostatic equilibrium, but for the
purposes of this illustration, we assume that the models do stay in
hydrostatic equilibrium and calculate the time between models at
$\Delta t = \Delta E/(L_{\rm in-L_|rm out})$, where $\Delta E$ is the energy
difference between models.
These calculations are described in more detail in Sect.~\ref{evtime}.}
\end{figure}

As discussed above,
if the luminosity at the base is strongly super-critical,
then the envelope will expand and the photosphere will move to larger
radii (see Fig.~2).  For an envelope with its photosphere below
$\sim 10^{14}$~cm, the dynamical time is shorter than the thermal
time, so the envelope should expand in hydrostatic equilibrium.  When
the photosphere grows larger than $\sim 10^{14}$~cm, the envelope can
no longer be in hydrostatic equilibrium. However, the material at
small radii has a much shorter dynamical time, so may be able to
adjust to a near-equilibrium state quickly. The envelope
could continue to expand, perhaps with the interior in a near
equilibrium state and the outer parts further from equilibrium.  If
the photospheric radius reaches $\sim 10^{15}$~cm, the thermal time
scale is so short and the pressure scale height becomes so large
compared to the radius, that it seems most likely either a strong wind
would form and carry the excess energy away, or the entire envelope
would be unbound.

Fig.~2 illustrates possible evolutionary
sequences for different values of the base luminosity.  For Fig.~2,
we make the assumption that the envelope is in a hydrostatic state
(described by eqn.~\ref{eq8}), even after the thermal time scale falls
below the dynamical time scale. The time to move between envelopes
$\Delta t = \Delta E/L_{+}$, where $\Delta E$ is the energy difference
between sequential envelopes (A similar method was applied by Ulmer
(1998b) to investigate the evolution of thick accretion disks.)
Evolutionary sequences show the evolution of both photospheric radius
and temperature with time.  The curves become nearly vertical as the
envelopes evolve to large radii, because the binding energy becomes so
small that they can be unbound by very little energy input.

It seems likely that the envelope will expand on a time scale of months
to years, depending on the base radius and the base luminosity, to
$\sim 10^{14}$~cm where $T_{\rm eff} \sim 50,000$~K. Subsequently, the 
envelope will expand, but will not be able to maintain hydrostatic equilibrium
throughout.

\section{Discussion \label{efinsec}}

Eddington envelopes require fine tuning
of the luminosity in order to produce extended envelopes.
It appears unlikely that the luminosity is held within such a tight region.
More reasonably, either the average luminosity will be sub-critical
at the base, and the envelope will collapse, or the luminosity will
be super-critical at the base, and the envelope will grow (and thereby
radiate at lower temperatures between $\sim$80,000~K and $\sim$20,000~K
as shown in Fig.~2) and eventually reach the most extended hydrostatic
model or the wind solutions, as described in Sect.~\ref{clossec}.
This scenario may be valid only if convection is efficient.
The stellar structure equations are not as tractable if
convection is inefficient, and we have not attempted to solve them here.

We reiterate that the remarkable feature
of extended envelopes is that they could reprocess much of
the hard radiation produced by a tidal disruption event into
the optical band.
Bolometric corrections would drop from $\sim 7.5$~magnitudes to 1--2.
The color temperatures of the objects may be slightly higher than the
effective temperatures due to suppression of absorption processes
at such low densities, but such changes are expected to be about
a factor of two
(see the work on neutron stars by Babul \& Paczy\'nski (1987)).
Even scattering-dominated envelopes retain their important property
of producing optically bright flares.

\begin{acknowledgements}
We thank A. Loeb and H.~C. Spruit for comments on the manuscript.
AU was supported in part by an National Science Foundation
graduate fellowship. We acknowledge
NSF grant AST9530478.
\end{acknowledgements}

\end{document}